\documentclass[useAMS,usenatbib]{mn2e}
\usepackage{graphicx}
\usepackage{lscape}

\title[Sample of 1-2 M$_\odot$ eclipsing binaries]
{Orbital and physical parameters of eclipsing binaries from the 
ASAS catalogue --- I. A sample of systems with components' masses 
between 1 and 2 M$_\odot$}
\author[K. G. He{\l}miniak, M. Konacki, M. Ratajczak and M. W. Muterspaugh]
{K. G. He{\l}miniak$^{1}$\thanks{E-mail:xysiek@ncac.torun.pl},
M. Konacki$^{1,2}$, M. Ratajczak$^{1}$ and M. Muterspaugh$^{3,4}$
\\
$^{1}$Nicolaus Copernicus Astronomical Center, 
Department of Astrophysics, ul. Rabia\'{n}ska 8 , 
87-100 Toru\'{n}, Poland\\
$^{2}$Astronomical Observatory, A. Mickiewicz University, ul. S{\l}oneczna 36, 
60-286 Pozna\'{n}, Poland\\
$^{3}$Department of Mathematics and Physics, College of Arts
and Sciences, Tennessee State University, Boswell Science Hall,
Nashville, \\TN 37209, USA\\
$^{4}$Tennessee State University, Center of Excellence in Information
Systems, 3500 John A. Merritt Blvd., Box No. 9501,
Nashville, \\TN 37203-3401, USA\\
}

\begin{document}

\date{Accepted ... ; Received ... ; in original form .... }

\pagerange{\pageref{firstpage}--\pageref{lastpage}} \pubyear{2009}

\maketitle

\label{firstpage}

\begin{abstract}
We derive the absolute physical and orbital parameters for a sample of 18
detached eclipsing binaries from the \emph{All Sky Automated Survey}
(ASAS) database based on the available photometry and our own radial velocity
measurements. The radial velocities (RVs) are computed using spectra
we collected with the 3.9-m Anglo-Australian Telescope and its \emph
{University College London Echelle Spectrograph} and the 1.9-m SAAO Radcliffe telescope and its
\emph{Grating Instrument for Radiation Analysis with a Fibre Fed Echelle}. 
In order to obtain as precise RVs as possible, most of the systems were 
observed with an iodine cell available at the AAT/UCLES and/or analyzed
using the two-dimensional cross-correlation technique (TODCOR). The RVs were
measured with TODCOR using synthetic template spectra as references. However, 
for two objects we used our own approach to the tomographic disentangling
of the binary spectra to provide observed template spectra for the
RV measurements and to improve the RV precision even more. For one
of these binaries, AI~Phe, we were able to the obtain an orbital solution 
with an RV $rms$ of 62 and 24 m s$^{-1}$ for the primary and secondary
respectively. For this system, the precision in $M \sin^3{i}$ is 0.08\%.

For the analysis, we used the photometry available in the ASAS database. We combined 
the RV and light curves using PHOEBE and JKTEBOP codes to obtain the absolute 
physical parameters of the systems. Having precise RVs we were able to reach 
$\sim$0.2 \% precision (or better) in masses in several cases but in radii, due to the limited 
precision of the ASAS photometry, we were able to reach a precision of only 1\% in 
one case and 3-5 \% in a few more cases. For the majority of our objects, the orbital 
and physical analysis is presented for the first time. 
\end{abstract}


\section{Introduction}
The importance of eclipsing binaries for modern astrophysics can not be 
underestimated. Their favorable geometry enables one to measure many basic 
physical parameters of the components, like masses, radii or luminosities
which are crucial for our still not complete understanding of stellar evolution. 
Eclipsing binaries with components in various evolutionary stages 
work as benchmarks for theory of evolution, distance indication, chemical 
and dynamical history of the Galaxy. Also a hot-topic in modern astronomy 
-- exoplanets -- relies on the knowledge of the host star, its mass and
radius. Large, automated, photometric surveys open new possibilities 
in eclipsing binaries' characterization. Thousands of light curves of eclipsing 
binaries are being produced for which $\sim$10 m$\,$s$^{-1}$ and better radial velocity
precision can be obtained \citep{Konacki:09::} and enable one to derive stellar 
parameters with a precision easily below 1.0\%. In particular the precision
in masses may reach 0.01-0.1\% level.

In this paper we present the first results of our on-going spectroscopic survey of the 
eclipsing binaries from the {\it All Sky Automated Survey}
\citep{poj02,pacz06}. Our 
spectroscopic follow-up to provide radial velocities (RVs) was carried out with 
two high-resolution echelle spectrographs: the \emph{University College London Echelle
Spectrograph} (UCLES) at the 3.9-m Anglo-Australian Telescope (AAT) and the 
\emph{Grating Instrument for Radiation Analysis with a Fibre Fed Echelle}
(GIRAFFE) at the the 1.9-m SAAO Radcliffe telescope. At the AAT/UCLES we were 
able to use the iodine cell to improve the radial velocity
precision and in consequence the precision in masses of the binary stars'
components down to a level better than 0.5\%. The available ASAS photometry, 
when combined with our RVs, enables us to obtain binary star models
parameterized with the absolute values of their orbital and physical parameters. 
The spectroscopic observations, data reduction and the best-fitting
orbital/physical solutions for 18 detached eclipsing binaries (DEBs) from the ASAS
database are described below. 

\section{Objects}
\begin{table*}
\caption{Eighteen detached eclipsing binaries from the ASAS database and 
our spectroscopic survey. In the 'Telescope/Spectrograph' 
column 'R/G' denotes 'Radcliffe/GIRAFFE' and 'A/U' denotes 'AAT/UCLES'. If the system 
was not previously known as an eclipsing or spectroscopic binary before ASAS, a 'Yes' 
flag is given in the column 'New?'. If it was, a reference is given.
\label{tab_info}}
\begin{tabular}{ccccrrrrll}
\hline
ASAS ID & Other ID & Tel./Sp. & Dist [pc] & V [mag] & J [mag] & 
H [mag] & K [mag] & Sp.T. & New? \\
\hline \hline
010538-8003.7&	CD-80~28	& R/G & --- & 10.10 & 8.49 & 8.04 & 7.83 & K2 & Yes \\
010934-4615.9$^a$&AI~Phe, HD~6980& A/U& 256 &  8.60 & 7.30 & 6.94 & 6.82 & G8 & No (1)\\
014616-0806.8&	BD-08~308	& A/U & --- & 10.52 & 9.43 & 9.15 & 9.05 & G2 & Yes \\
023631+1208.6&	BD+11~359	& A/U & --- &  9.75 & 8.70 & 8.51 & 8.42 & F7 & Yes \\
042041-0144.4&	HIP~20267	& A/U & 116 &  8.70 & 7.53 & 7.25 & 7.19 & G3.5 & Yes \\
042724-2756.2$^b$& IDS04234-2810~A & A/U &---& 9.89 & 8.98 & 8.84 & 8.71 & F6 & Yes\\
053003-7614.9$^{ab}$& UX~Men, HD~37513& A/U& 101& 8.23 & 7.20 & 6.98 & 6.92 & F7 & No (2)\\
071626+0548.8&	TYC~176-2950-1	& A/U & --- & 10.19 & 8.97 & 8.65 & 8.57 & G6 & Yes \\
085524-4411.3&	CD-43~4765	& A/U & --- & 10.04 & 9.00 & 8.80 & 8.76 & F7 & Yes \\
150145-5242.2&	HD~132553	& A/U & --- &  9.58 & 8.46 & 8.20 & 8.19 & G0 & Yes \\
155259-6637.8&	HD~141344	& A/U & --- &  8.99 & 7.97 & 7.75 & 7.69 & F7 & Yes \\
155358-5553.4&	HD~141857	& A/U & --- &  9.56 & 8.45 & 8.28 & 8.15 & G0 & Yes \\
162637-5042.8$^b$& HD~147827	& A/U & --- &  9.81 & 8.69 & 8.44 & 8.37 & G1 & Yes \\
174626-1153.0&	BD-11~4457 	& R/G & --- & 10.81 & 8.59 & 8.16 & 8.07 & K4.5 & Yes \\
185512-0333.8&	HD~175375	& A/U & --- & 10.16 & 9.12 & 8.97 & 8.92 & F6 & Yes \\
193044+1340.3& V415~Aql, HD~231630& A/U& 910& 10.41 & 9.55 & 9.27 & 9.18 & F6 & No (3)\\
195113-2030.2&	HD~187533	& A/U & --- &  9.77 & 8.63 & 8.41 & 8.34 & G1 & Yes \\
213429-0704.6&	BD-07~5586	& A/U & --- & 10.53 & 9.47 & 9.24 & 9.15 & G0 & Yes \\
\hline
\end{tabular}
\smallskip
\flushleft
$^a$ Photometry is available from ASAS but the star is not indicated as a variable.\\
$^b$ Star has a visual and/or physical companion.\\
Note: The available distances are based on the {\it Hipparcos} parallaxes \citep{per97}, 
except for V415~Aql \citep{bra80}. $V$ is the maximum magnitude from ASAS, 
$J, H$ and $K$ are taken from 2MASS \citep{cut03}. The spectral type is based on
$T_{eff}$ vs. $V-K$ relation for dwarfs by \citet{tok00}.\\
Ref.: (1) \citealt{and88}; (2) \citealt{and89}; (3) \citealt{bra80}.\\
\end{table*}

The targets of this spectroscopic survey are detached spectroscopic binaries
with spectral types later than $\approx$F5 for which precise RV
measurements can be made. In order to select the appropriate targets, we 
proceeded as follows. The {\it ASAS Catalog of Variable Stars} \citep[ACVS;][]{poj02}
was searched for DEBs with with no obvious out-of-eclipse variations, 
possibly short-lasting eclipses and with $V-K>1.1$. There are 16 such
stars in this paper. Two other systems ASAS J010538-8003.7 ($V-K=2.27$) and 
ASAS J174626-1153.0 ($V-K=2.74$) are from our separate observing program.
In order to select relatively bright objects and limit the exposure 
times but to still have a relatively large sample, we searched for binaries
with $V\le11\,$mag. 

In Table \ref{tab_info} we present the basic characteristics of the targets 
discussed below. Fifteen stars turned out to be new variables whose eclipsing 
nature was first reported in the ACVS. All the systems have their ASAS photometry
available on-line\footnote{http://www.astrouw.edu.pl/asas/?page=main}. The
time-span of the ASAS photometry exceeds 8 years, hence a good phase coverage
and accurate period determination may be expected. However, the two
brightest targets in our sample are not indicated in the ACVS as variables. 
These are AI~Phe (ASAS J010934-4615.9) and UX~Men (ASAS J053003-7614.9) 
previously characterized by \citet{and88} and \citet{and89} respectively. 
For AI~Phe, our novel implementation of the iodine cell 
technique for spectroscopic binaries \citep{Konacki:09::} allows us to improve 
its parameters, especially masses, to an unprecedented level of precision. 
For UX~Men we have also obtained a higher precision in masses even though this 
system has wider spectral lines than AI~Phe which reduces the attainable RV 
precision. We have also substantially improved the characteristics of the third 
of the previously known eclipsing binaries in this sample -- V415~Aql 
\citep[ASAS J193044+1340.3;][]{bra80} -- which until now was reported to have 
2 times shorter period, a large brightness ratio and no secondary eclipse.

\section{Observations}

\subsection{Radcliffe/GIRAFFE}
Spectra of the systems ASAS J010538 and ASAS J174626 
were obtained during two runs in June (ASAS J010538) and October 2006 (ASAS J174626) 
with the 1.9-m Radcliffe telescope and GIRAFFE as a part of our 
low-mass eclipsing binaries search program. 
GIRAFFE provides spectra with a resolution of $\simeq 40000$. Due to a
relatively low throughput of the entire system, we used the exposure
time of $3600\,$s. The resulting signal-to-noise ratio (SNR) per collapsed
spectral pixel varied and depending on the observing conditions was $\sim$35 to 
$\sim$70 for both objects. The wavelength calibration was done in a standard 
manner with a ThAr lamp exposure taken before and after a stellar exposure.

\subsection{AAT/UCLES}
The rest of the objects were observed with the AAT/UCLES during 3 runs  
(11 nights) between September 2008 and January 2009. We used a 1" slit which
provides a resolution of $\simeq 60000$. Most of the time
we adopted an exposure time of $900\,$s. In good seeing conditions we 
were able to obtain an $SNR \sim$90 for our typical target and
an exposure without the iodine cell. However, usually the SNR was between 30 
and 65. In bad seeing conditions we obtained a $SNR \sim$30-40 
for the brightest targets and no iodine cell in the light path. 
If weather permitted we used an iodine cell. The exposures
with the iodine cell had a $SNR$ about 30\% lower than without the cell. 
The ThAr lamp exposures were also taken throughout each night but not 
after every single stellar exposure.

An iodine cell becomes useful when a $SNR$ is $\sim$50 or more. 
However, spectra with a $SNR$ as low as $\sim$30 taken through an
iodine cell can still be reduced. For most of the nights at the AAT
we had to deal with large (above 2") and variable seeing and it was
not always possible to decide beforehand if it was practical to use
the iodine cell for a given target. In consequence, we ended up with 
a number of exposures taken through the iodine cell with an $SNR$ too low
for high RV precision. Fortunately, since the iodine cell
approach for binary stars requires always taking pairs of exposures
with and without the cell \citep{Konacki:09::,Konacki:05::}, if
we could not use or take an exposure with the cell we always had one
without the cell for each target. These were subsequently used to
measure an RV with the usual ThAr based approach. In consequence,
we have three types of RV datasets --- based entirely on the
iodine cell, entirely on the ThAr wavelength calibration or
mixed sets when both types of calibrations are used to
provide RVs.

\section{Analysis}
\subsection{Radial Velocities}
The raw ccd frames taken on both telescopes/spectrographs were reduced in a 
standard manner (bias subtraction, flat fielding) with {\it IRAF} 
package {\it ccdred}\footnote{{\it IRAF} is written and supported by the {\it IRAF}
programming group at the National Optical Astronomy Observatories (NOAO)
in Tucson, AZ. NOAO is operated by the Association of Universities for 
Research in Astronomy (AURA), Inc. under cooperative agreement with the
National Science Foundation. http://iraf.noao.edu/}. The subsequent 
echelle data reduction was also carried out with {\it IRAF} and 
its {\it echelle} package. The RVs from the ThAr wavelength calibrated 
spectra were calculated with our own implementation of the two dimensional
cross-correlation technique \citep[TODCOR]{zuc94}. As templates, we used 
synthetic spectra computed with the ATLAS9 and ATLAS12 codes \citep{kur92}.
With the exception of two systems, AI~Phe (ASAS J010934) and UX~Men
(ASAS J053003), the RVs from the iodine cell calibrated spectra were also 
computed with TODCOR. The details of such a procedure are described by 
\cite{Konacki:05::}.

For the two brightest binaries in this sample, UX~Men (ASAS J053003) 
and AI~Phe (ASAS J010934) we were able to collect eight spectra
taken with the iodine cell each. This is about the smallest number
of spectra still sufficient to carry out a tomographic disentangling
to obtain observed component spectra of a binary. Our disentangling 
procedure is described by \cite{Konacki:09::}. It essentially allows one to
derive the component spectra from the observed composite spectra and 
then use them to compute the RVs. As we have shown this approach 
is capable of providing RVs of the
components of double-line binary stars with a precision reaching
5 m$\,$s$^{-1}$ \citep{Konacki:09::}. While the spectra of UX Men
and AI Phe from the AAT/UCLES are characterized by a much lower 
$SNR$ of $\sim$40-100 compared to those used in \cite{Konacki:09::}, 
the disentangling still can be carried out. This procedure has resulted 
in a higher RV precision compared to the standard ThAr approach or the 
iodine cell based approach combined with TODCOR for these two targets. 
In particular, the best-fitting RV solution for AI~Phe is characterized by an $rms$
of 62 and 24  m$\,$s$^{-1}$ for the primary and secondary respectively.
The UX~Men $rms$ is not nearly as good (210 and 270 m$\,$s$^{-1}$) but 
this is due to the very wide spectral lines of its components. There
is no doubt that the RV precision for AI Phe would be much higher
if higher $SNR$ spectra were available. It should be noted here
that the RV precision from the iodine cell spectra used in
this paper is not representative for this technique as most
of the time we were dealing with an $SNR$ far too low for what
is required to obtain a high RV precision. Still the results
are quite satisfactory precision-wise.

For seven systems the iodine cell based solution was substantially better 
than the ThAr based one. For four other targets we used RVs based on both methods
of wavelength calibration since the number of iodine cell based spectra 
was too small or the resulting $rms$ was comparable. The seven remaining 
targets (including the two observed at SAAO) have their solutions based 
on the ThAr calibrated spectra only.

The RV measurements together with their errors and $O-C$ are collected 
in Table \ref{tab_allrv} in the Appendix \ref{sec_allrv}. 
As it turned out the formal errors $\sigma_0$ computed from the scatter
between the echelle orders used in the analysis were somewhat
underestimated. Hence to obtain $\chi^2 \simeq 1$ for our RV solutions
and more conservative estimates of the errors of the best-fitting orbital
and physical parameters, we added in quadrature an additional error 
$\sigma_{sys}$ to the formal RV errors. For every component of every binary the 
additional RV error was estimated independently and in general varied 
from star to star. There are a few possible sources of such errors.
For the ThAr based RVs, the largest additional contribution to the error 
budget comes from a wavelength solution based on ThAr taken sometime
before or after a stellar exposure and hence usually not just before
or after. For the iodine cell based RVs, most likely the additional RV 
error comes from an imperfect modelling of the spectrograph's point spread 
function imposed by a relatively low SNR of the spectra. Presumably,
also the intrinsic RV variability of the stars itself contributes to
the error budget. In Table \ref{tab_allrv} final RV errors are shown.

\subsection{Modeling}

\begin{table*}
\caption{Absolute orbital and physical parameters from our solutions for 
the investigated targets with $1\sigma$ uncertainties in parentheses.}
\label{tab_master}
\begin{tabular}{ccccccccc}
\hline
ASAS ID & Solution & $P$ & $T_0$ (PHOEBE) & $K_1$ & $K_2$ & $v_\gamma$ & $q$ & $a$  \\
& & [d] & JD-2 450 000 & [km s$^{-1}$] & [km s$^{-1}$] & [km s$^{-1}$] & & [R$_{\sun}$] \\
\hline\hline
010538-8003.7 &th-ar & 8.069406(6) &  1873.449(6) & 73.6(1.6)& 73.3(2.4)&  -5.8(7)  & 1.003(31) & 23.4(5) \\
010934-4615.9 &iodine$^a$&24.59241(8)& 3247.184(3)& 51.16(3) & 49.11(2) & -0.750(12)& 1.0418(8) & 47.855(19) \\
014616-0806.8 &mixed & 5.940047(25)&  1878.427(7) & 76.2(1.1)& 84.2(8)  &  35.0(6)  & 0.905(16) & 18.79(16) \\
023631+1208.6 &th-ar & 3.604913(8) &  2449.258(2) & 85.8(4)  &102.3(3)  &   6.2(2)  & 0.839(7)  & 13.40(4) \\
042041-0144.4 &iodine& 6.47649(2)  &  1943.030(6) & 74.27(38)& 81.48(38)& -17.2(1)  & 0.911(7)  & 19.76(7) \\
042724-2756.2 &iodine& 8.94657(5)  &  1873.707(7) & 74.0(4)  & 70.7(4)  &   7.79(4) & 1.047(8)  & 25.6(1) \\
053003-7614.9 &iodine$^a$&4.181096(3)&2019.2755(11)&87.31(12)& 89.90(8) &  48.86(5) & 0.9712(16)& 14.639(12) \\
071626+0548.8 &mixed & 11.55478(6) &  2426.025(8) & 61.2(3)  & 62.1(3)  &  15.6(3)  & 0.986(7)  & 27.5(1) \\
085524-4411.3 &mixed & 7.040274(14)&  1875.092(4) & 68.0(1)  & 77.50(12)&  21.23(7) & 0.8772(19)& 20.279(22) \\
150145-5242.2 &iodine& 5.976930(76)&  1927.389(7) & 89.1(7)  & 88.9(7)  &   4.6(7)  & 1.002(10) & 21.01(11) \\
155259-6637.8 &iodine& 5.744754(22)&  1931.664(6) & 93.47(12)& 76.98(6) &  11.00(5) & 1.214(2)  & 19.367(15) \\
155358-5553.4 &iodine& 5.691743(23)&  1933.457(6) & 94.59(5) & 90.67(9) &   4.83(4) & 1.043(1)  & 20.877(11) \\
162637-5042.8 &th-ar & 8.87621(5)  &  1939.350(11)& 65.67(17)& 77.39(35)& -28.5(1)  & 0.852(2)  & 25.16(5) \\
174626-1153.0 &th-ar & 3.011055(7) &  1956.245(3) &116.2(9)  & 96.2(1.5)& -57.2(7)  & 1.21(2)   & 12.63(11) \\
185512-0333.8 &th-ar & 5.795527(15)&  1972.545(4) & 80.40(5) & 96.37(10)& -17.08(5) & 0.834(1)  & 20.257(13)\\
193044+1340.3 &th-ar & 4.925489(16)&  2731.560(4) & 91.34(55)& 98.0(6)  &  13.5(5)  & 0.93(1)	& 18.44(8) \\
195113-2030.2 &mixed & 7.04335(5)  &  1976.202(9) & 65.41(14)& 90.71(21)&  -8.1(1)  & 0.7211(23)& 21.873(35) \\
213429-0704.6 &th-ar & 5.672517(8) & 1884.2545(25)& 75.(2)   & 79.3(7)  & -21.0(7)  & 0.95(3)   & 17.24(24) \\
\hline
ASAS ID& $e$ & $\omega$ & $i$ & $M_1$ & $M_2$ & $R_1$ & $R_2$ & $T2/T1$ \\  
& & [$^\circ$] & [$^\circ$] & [M$_{\sun}$] & [M$_{\sun}$] & [R$_{\sun}$] & [R$_{\sun}$] & \\
\hline \hline
010538-8003.7 & 0.0	 &    ---  & 80.4(6)  & 1.380(55) & 1.384(55) & 3.16(33) & 4.06(41) & 0.72(3)\\
010934-4615.9 & 0.187(4) & 110.1(9)& 84.4(5)  & 1.2095(11)& 1.2600(11)& 1.82(5)  & 2.81(7)  & 0.820(15)\\
014616-0806.8 & 0.067(45)& 100(5)  & 86.1(7)  & 1.335(25) & 1.208(24) & 1.57(53) & 1.42(52) & 1.014(80)\\
023631+1208.6 & 0.0	 &    ---  & 87.4(9)  & 1.357(8)  & 1.138(7)  & 1.82(36) & 1.28(29) & 0.988(18)\\
042041-0144.4 & 0.132(5) & 342.2(9)& 88.5(3)  & 1.293(9)  & 1.179(9)  & 1.59(29) & 1.43(26) & 0.99(11)\\
042724-2756.2 & 0.012(4) & 238(24) & 86(1)    & 1.383(13) & 1.449(13) & 2.2(1)   & 2.16(8)  & 0.94(3)\\
053003-7614.9 & 0.0	 &    ---  & 89.86(15)& 1.2229(15)& 1.1878(15)& 1.321(36)& 1.285(37)& 0.992(18)\\  
071626+0548.8 & 0.21(5)  &  52(10) & 88.2(7)  & 1.055(8)  & 1.041(8)  & 1.65(3)  & 1.18(2)  & 0.988(10)\\
085524-4411.3 & 0.0	 &    ---  & 86.4(2)  & 1.2040(27)& 1.0562(25)& 1.87(2)  & 1.634(16)& 0.996(6)\\
150145-5242.2 & 0.0	 &    ---  & 84.7(3.4)& 1.766(21) & 1.769(21) & 2.86(14) & 2.81(14) & 1.004(27)\\  
155259-6637.8 & 0.0	 &    ---  & 84.5(8)  & 1.352(3)  & 1.644(4)  & 1.83(15) & 2.81(15) & 0.984(35)\\
155358-5553.4 & 0.0	 &    ---  & 86.3(9)  & 1.847(6)  & 1.927(6)  & 2.73(28) & 2.99(30) & 0.97(5)\\
162637-5042.8 & 0.019(10)& 341(2)  & 85.6(3)  & 1.469(8)  & 1.246(7)  & 2.55(33) & 2.25(33) & 0.98(4)\\
174626-1153.0 & 0.0	 &    ---  & 81.4(5)  & 1.44(3)   & 1.74(3)   & 1.99(12) & 2.64(17) & 0.95(3)\\
185512-0333.8 & 0.0108(3)& 45(2)   & 87.4(1.4)& 1.813(6)  & 1.513(5)  & 1.7(3)   & 1.5(4)   & 0.87(4)\\
193044+1340.3 & 0.0	 &    ---  & 87.6(7)  & 1.798(16) & 1.675(15) & 2.97(9)  & 2.34(8)  & 0.974(16)\\
195113-2030.2 & 0.0	 &    ---  & 83.4(3)  & 1.646(8)  & 1.187(6)  & 2.74(13) & 2.03(11) & 0.994(18)\\
213429-0704.6 & 0.0937(8)& 170(10) & 89(1)    & 1.10(3)   & 1.04(3)   & 1.27(23) & 1.24(22) & 0.96(3)\\  
\hline
\end{tabular}
\smallskip
\flushleft
$^a$ Spectra were tomographically disentangled\\
\end{table*}

\begin{figure*}
\includegraphics[width=0.975\textwidth]{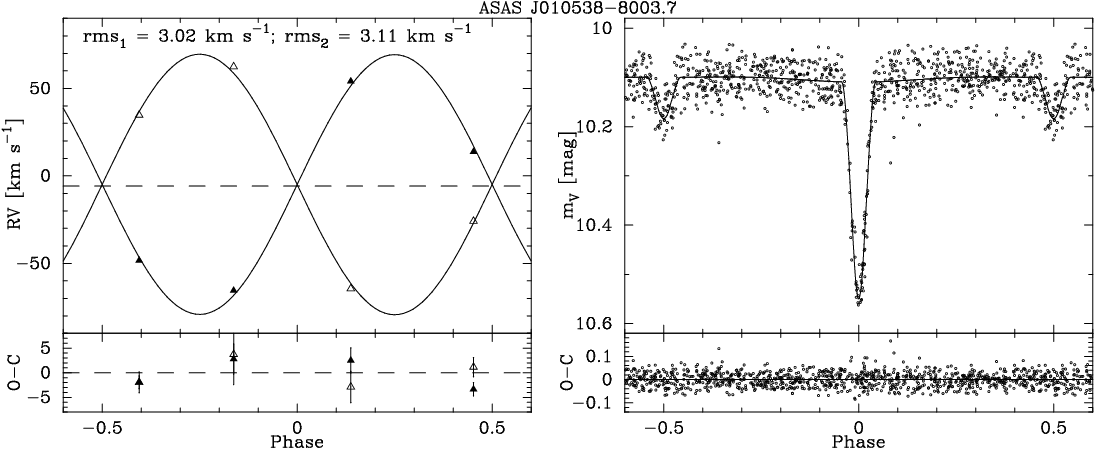}
\includegraphics[width=0.975\textwidth]{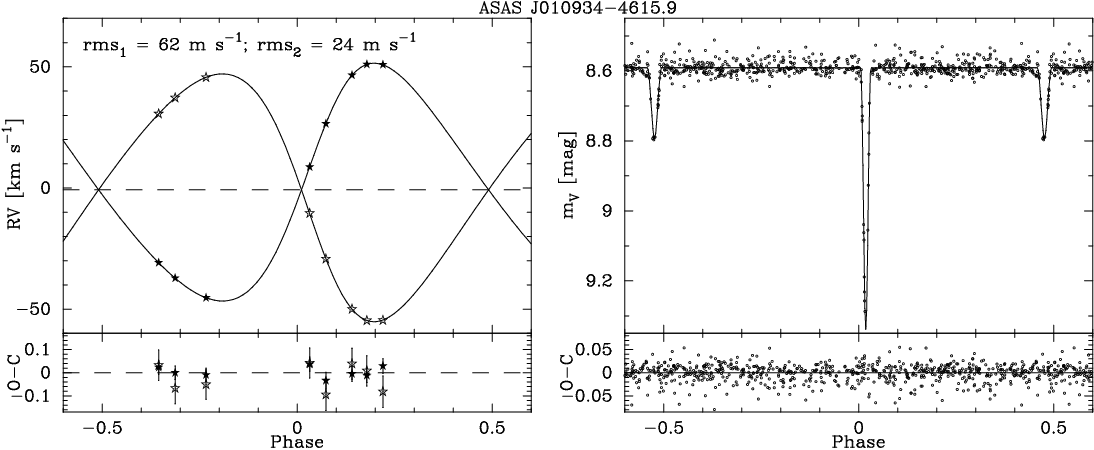}
\includegraphics[width=0.975\textwidth]{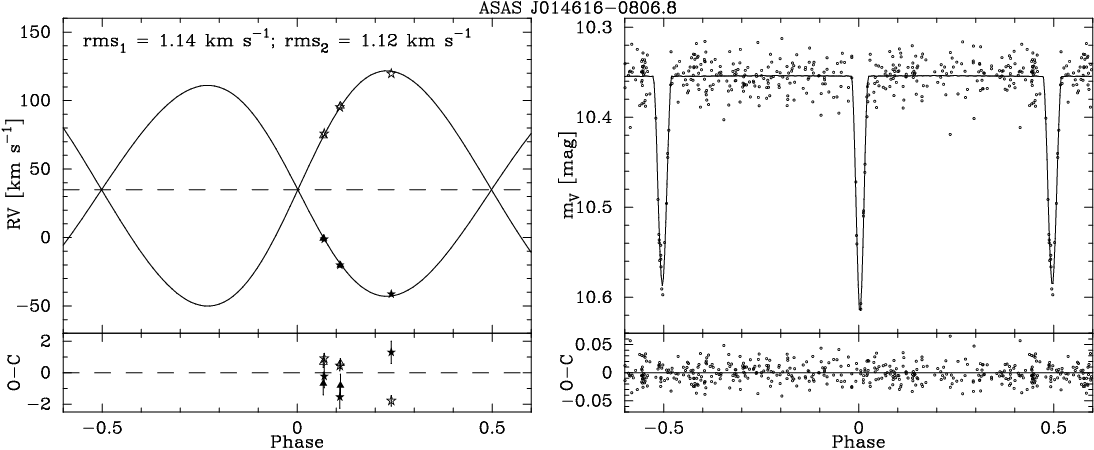}
\caption{Final solutions for all 18 systems. Plots are phase-folded with 
the correpsonding periods. Left panels show our RV measurements with 
the best-fitting solutions. Filled symbols are for the primaries
and open ones for the secondaries. Triangles are for the ThAr-calibrated and 
stars for the iodine cell-calibrated spectra. The resulting $rms$ and $O-C$ are
also shown.  Right panels depict the ASAS light curves and our model curves 
together with $O-C$.\label{fig_master}}
\end{figure*}
\begin{figure*}
\includegraphics[width=0.975\textwidth]{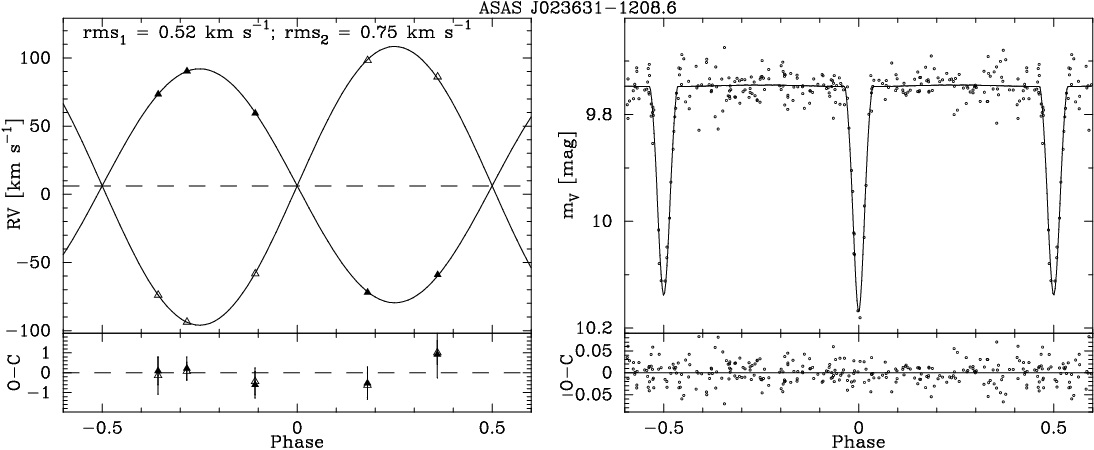}
\includegraphics[width=0.975\textwidth]{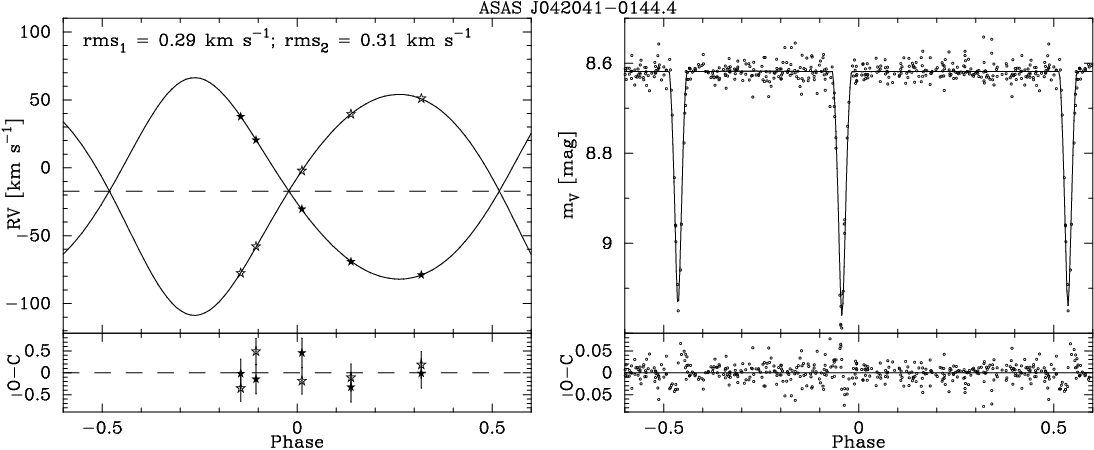}
\includegraphics[width=0.975\textwidth]{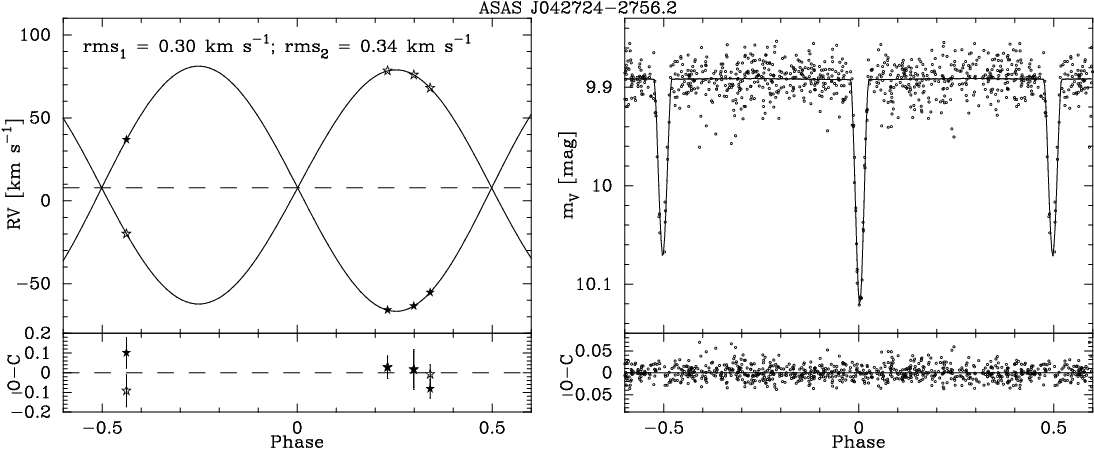}
\contcaption{}
\end{figure*}
\begin{figure*}
\includegraphics[width=0.975\textwidth]{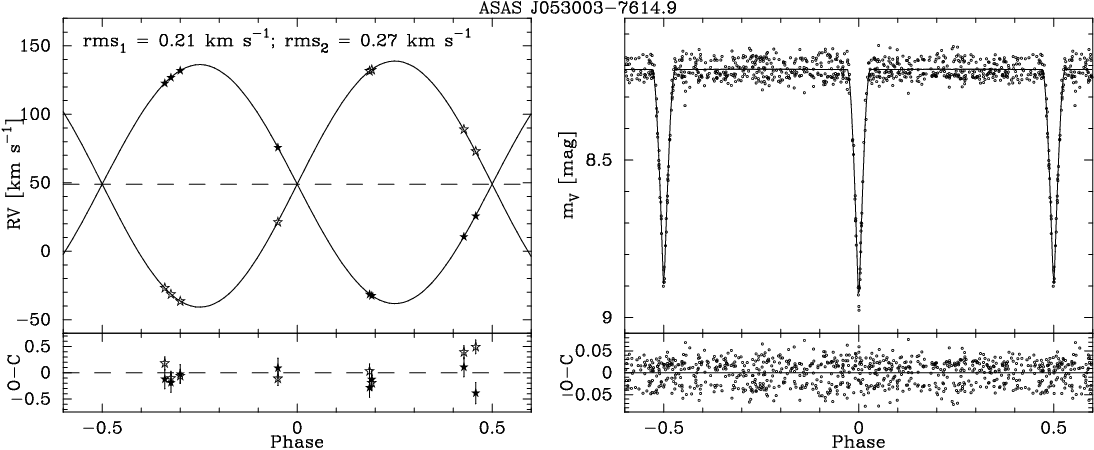}
\includegraphics[width=0.975\textwidth]{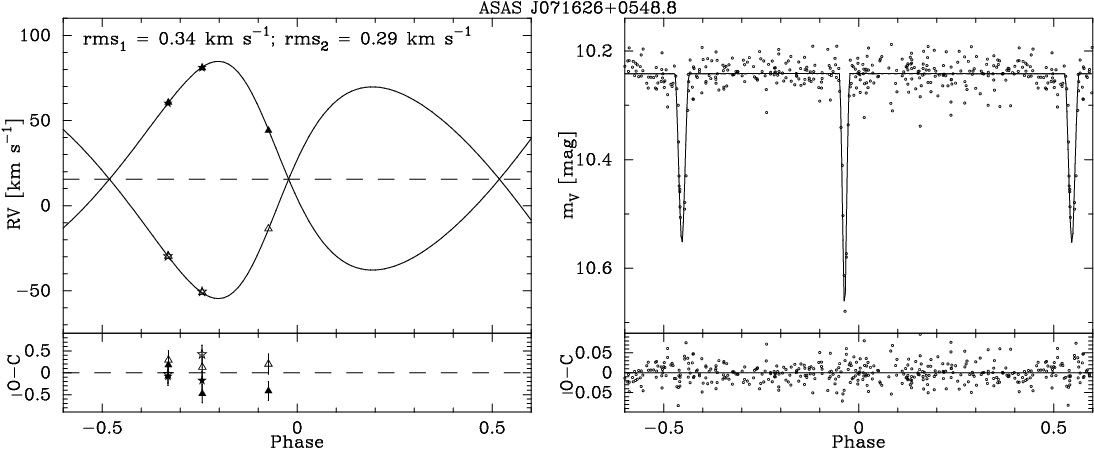}
\includegraphics[width=0.975\textwidth]{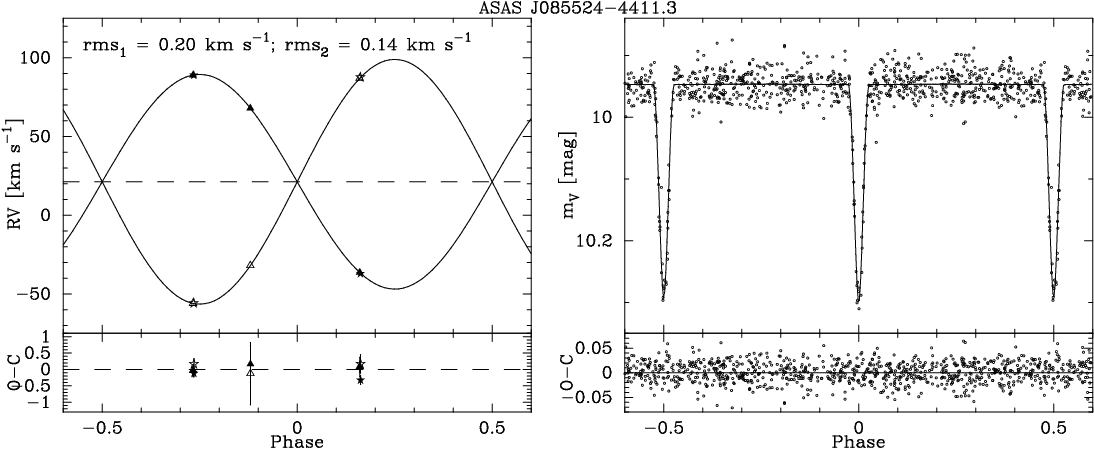}
\contcaption{}
\end{figure*}
\begin{figure*}
\includegraphics[width=0.975\textwidth]{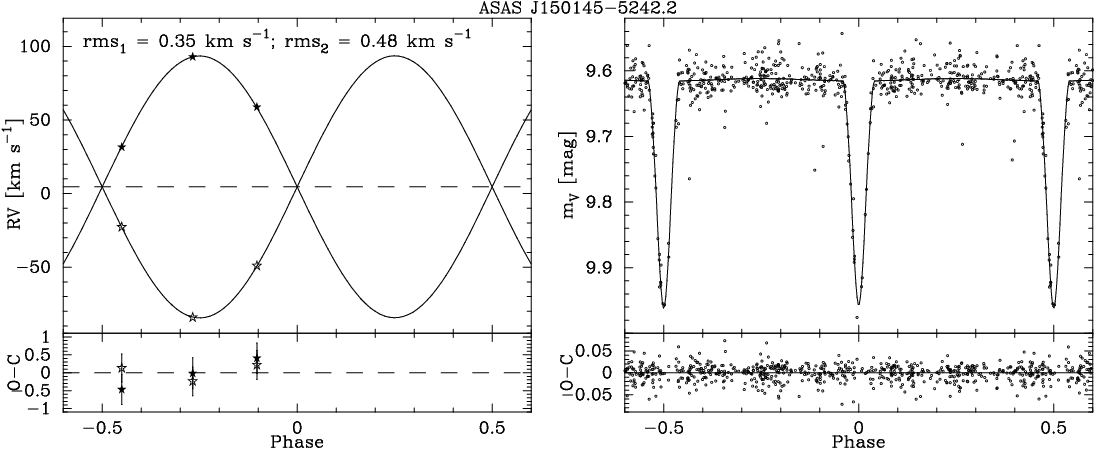}
\includegraphics[width=0.975\textwidth]{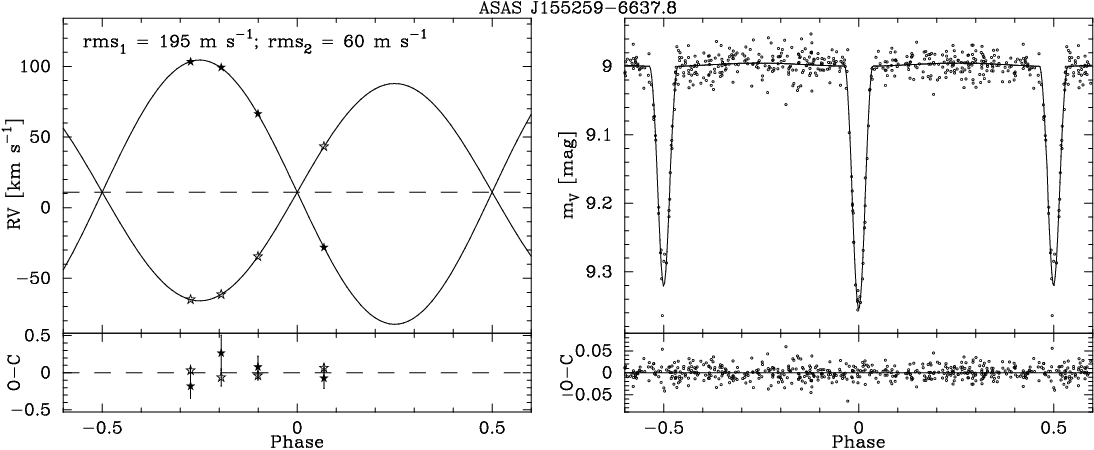}
\includegraphics[width=0.975\textwidth]{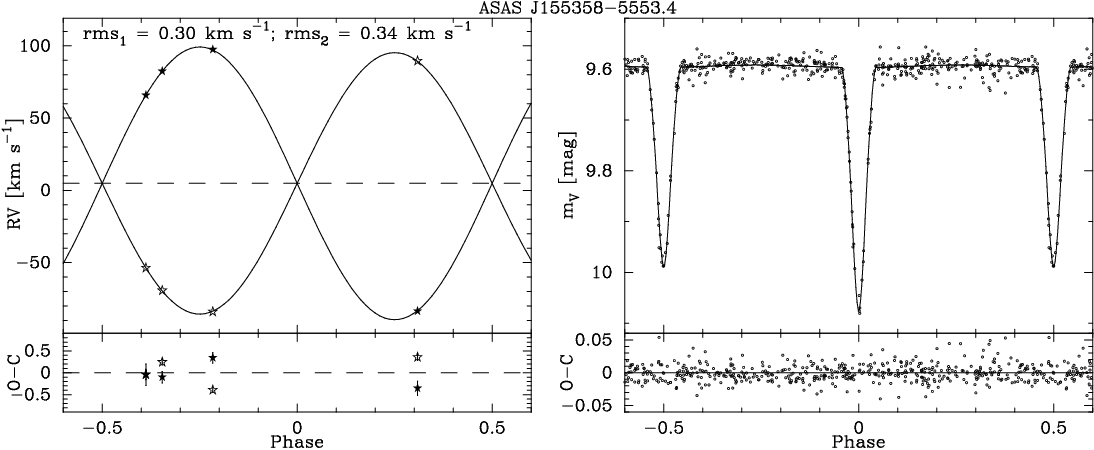}
\contcaption{}
\end{figure*}
\begin{figure*}
\includegraphics[width=0.975\textwidth]{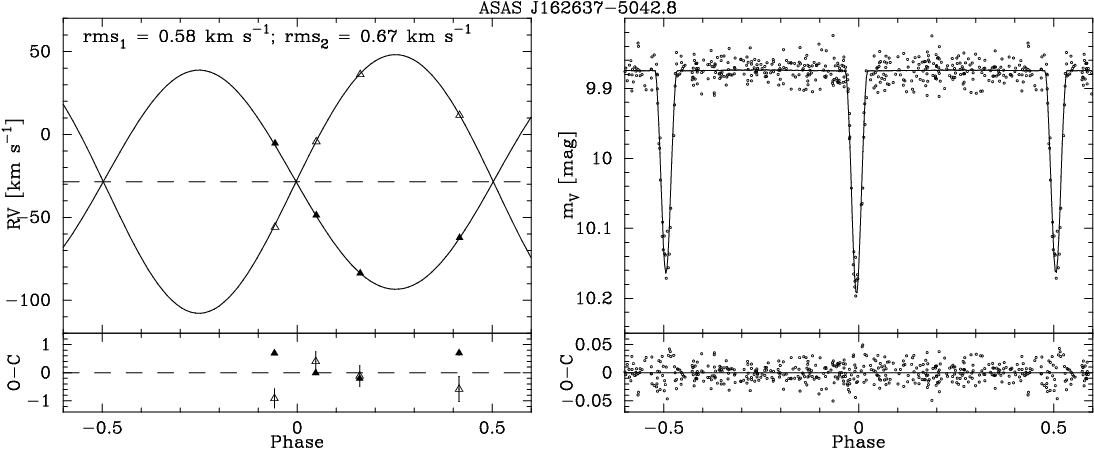}
\includegraphics[width=0.975\textwidth]{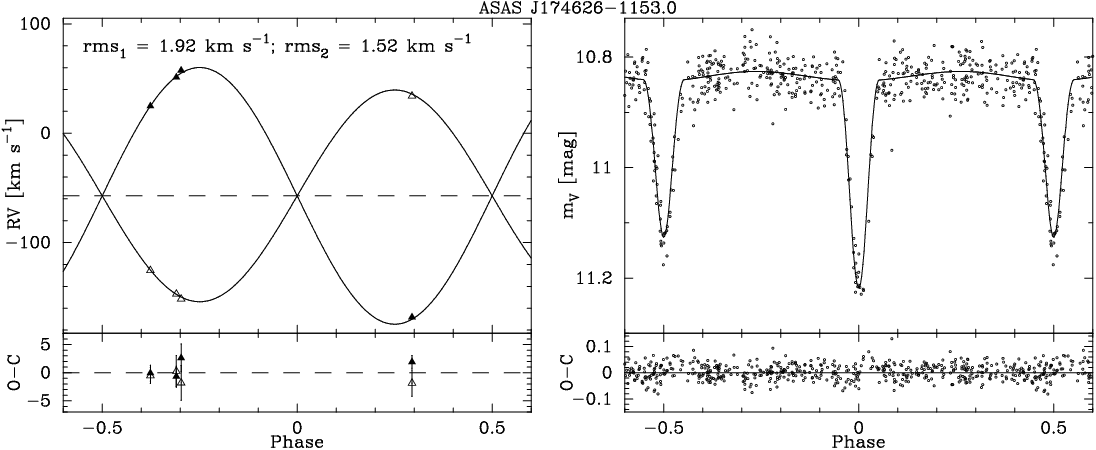}
\includegraphics[width=0.975\textwidth]{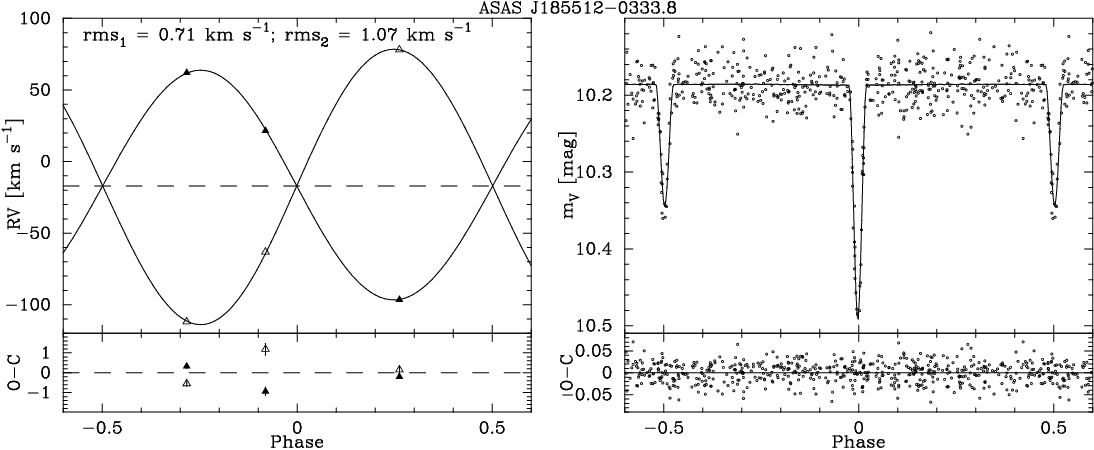}
\contcaption{}
\end{figure*}
\begin{figure*}
\includegraphics[width=0.975\textwidth]{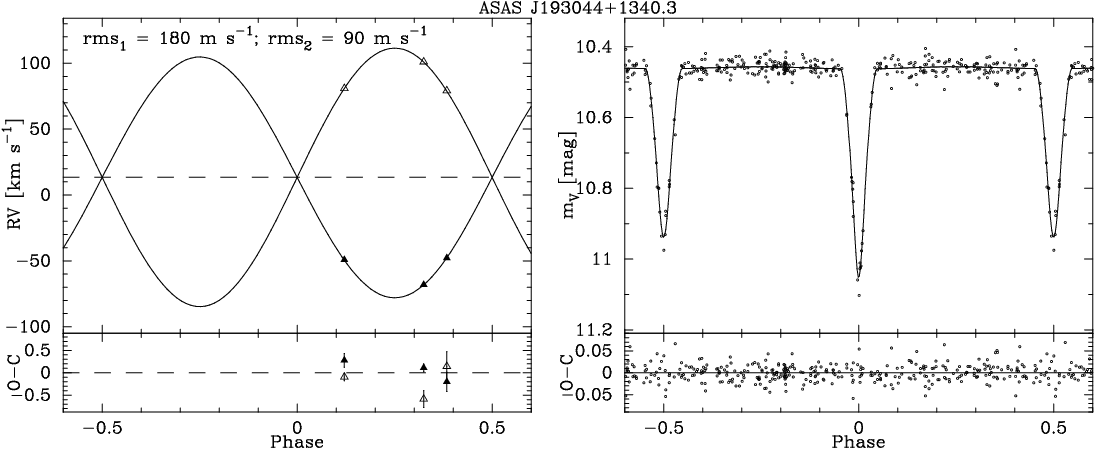}
\includegraphics[width=0.975\textwidth]{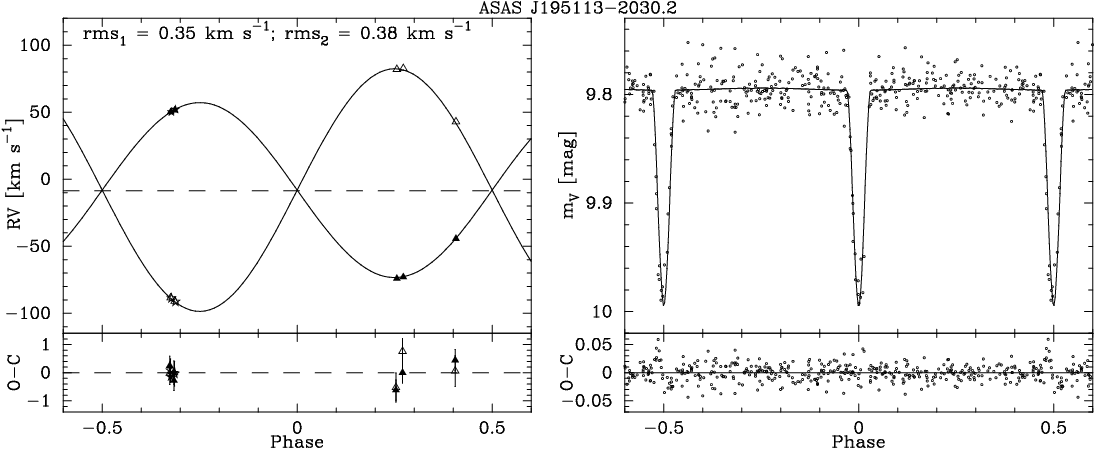}
\includegraphics[width=0.975\textwidth]{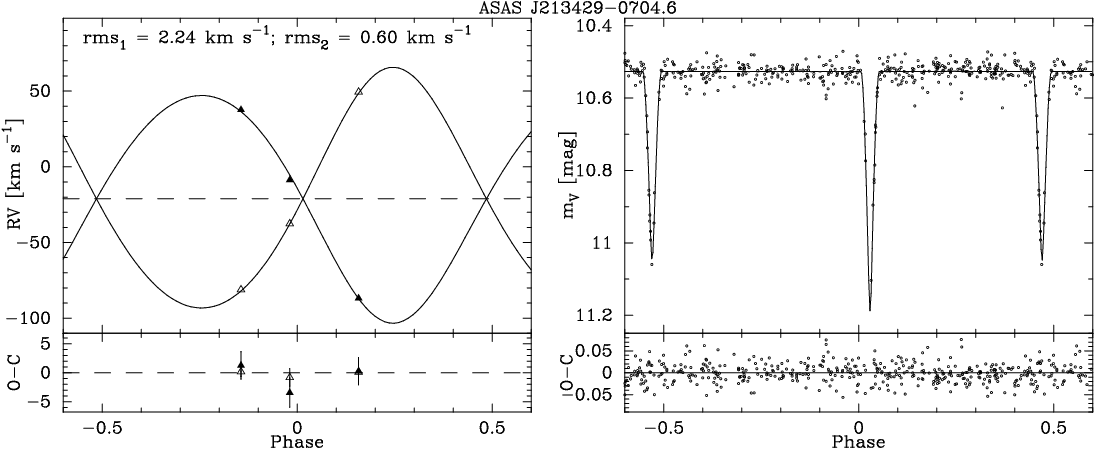}
\contcaption{}
\end{figure*}

Our radial velocity measurements were combined with the available ASAS
photometry to derive absolute physical and orbital parameters of 
our targets. We used three independent modeling codes, 
each of them for a different purpose, as follows.

First of all, we used the light curve modeling code JKTEBOP,
\citep{sou04a,sou04b} based on EBOP 
\citep[\emph{Eclipsing Binary Orbit Program};][]{pop81,etz81}, 
which fits a simple geometric model of a detached eclipsing 
binary to a single light curve. We obtained preliminary
yet quite accurate values of the orbital period $P$ and 
the moment $T_0$. A preliminary value of the mass ratio 
was assumed (usually close to or equal 1) in this step.

Second, we derived the orbital parameters using the RV 
measurements only. For that purpose we used our own software 
which fits a Keplerian orbit by minimizing the $\chi^2$ 
function with the least-squares Levenberg-Marquardt algorithm. 
With this software, for a circular orbit we fit for a zero 
phase (time of periastron for an eccentric orbit), two RV amplitudes 
and systemic velocity. For an eccentric orbit we also fit for an 
eccentricity and longitude of periastron. The fit is performed 
simulatenously to the RVs of both stars as by definition their 
longitudes of periastron differ by 180$^{\circ}$ but otherwise 
the stars follow the same orbit. This way from e.g. 5 visits to 
a binary we have 10 RV measurements and 4 (circular orbit) or 
6 (eccentric orbit) parameters to fit for. We held the
orbital period fixed at the value taken from JKTEBOP. 
The choice of a circular or eccentric RV orbit was based on 
a light curve only. This way we 
obtained precise values of the orbital/physical parameters related
to RVs, especially the mass ratio $q$ from the RV amplitudes $K_{1,2}$, 
which were later used in the next JKTEBOP runs. After the second 
JKTEBOP run we obtained improved 
values of $T_0$ and $P$ which were finally used with the third 
code -- {\it PHysics Of Eclipsing BinariEs} \citep[PHOEBE;][]{prs05},
an implementation of the Wilson-Devinney (WD) code \citep[with updates]{wil71}.

PHOEBE allows one to fit a model to the RV and light curves 
simultaneously. We used it to create the final model of each
binary. Our criterion for a best-fitting solution is a model given by 
PHOEBE which is also in agreement with the other two codes. However, 
the uncertainties given by PHOEBE tend to be underestimated so 
for the error estimations we used our RV-fitting code and 
JKTEBOP code that employs a Monte-Carlo simulator.

Binaries with eccentric orbits happen to cause a few problems
when analyzed in the way described. The main issue is the fact that 
every code defines its zero moment $T_0$ in a different way. 
In our RV-fitting code $T_0$ is the periastron passage. JKTEBOP 
defines $T_0$ as the moment of the deeper eclipse which is then called 
{\it primary} (eclipse of the {\it primary} star). In PHOEBE $T_0$ 
corresponds to the far intersection between the projection of the 
line-of-sight through the center of the ellipse on the orbital plane
and the ellipse \citep{prs06}. The primary eclipse in general 
{\it does not} coincide with $T_0$ and is defined as the one closer
to $T_0$. This convention is useful when an apsidal motion 
is present. In such a case no artificial period change is seen 
(as it is when $T_0$ is fixed to an eclipse). We decided to keep 
the convention from PHOEBE but to force the {\it deeper} eclipse to 
be closer to $T_0$. Hence we call the {\it primary} the star whose 
eclipse is the deeper one.

The initial values of the eccentricity $e$ and the periastron longitude
$\omega$ were calculated with JKTEBOP. Later, the values of $e$ and
$\omega$ were improved with our RV-fitting code. In the two eccentric
orbit cases when we only had three spectra and thus the number of RV 
measurements was only 6, $e$ and $\omega$ were not fitted with the RV 
code. The same was done for ASAS J014616 and ASAS J071626, despite 
having actually 5 measurements, as some of them (with and without 
I$_2$) were made during the same night one after another. The final values 
of $e$ and $\omega$ were determined with PHOEBE. This code is not particularly
efficient in fitting for $e$ and $\omega$ so we typically restarted
it with slightly different values of $e$ and $\omega$ to match
the difference in phase between the light curve's two minima as determined 
by JKTEBOP. Due to the different definitions of $T_0$, in some cases 
we had to change the initial value of $\omega$ by 180$^\circ$. 
The final set of the orbital parameters and the $O-C$ for the RVs 
and ASAS photometry are all in the framework of PHOEBE.

\section{Results}

The results of our modeling of 18 detached eclipsing binaries are shown 
in Table \ref{tab_master}. The orbital and physical parameters are 
given with their 1$\sigma$ uncertainties. We show the basic parameters
that are computed directly with the codes used or can be computed as a 
combination of the directly determined values. We do not present bolometric 
magnitudes and luminosities nor separate effective temperatures. 
The temperatures may be computed with PHOEBE but are highly dependent on 
their starting values. However, the temperature ratio is considered to be
reliable \citep{prs06} and it is shown in the last column of Tab. 
\ref{tab_master}. All our solutions are presented in Figure 
\ref{fig_master}. It should be noted that the reference phase ($\phi=0$) is 
defined as in PHOEBE. For eccentric orbits it means that this does not 
coincide with minima. For the purpose of ephemeris calculations and a possible 
further eclipse timing, in Table \ref{tab_min} we show the phases of the 
primary and secondary minima for 8 eccentric systems in our sample.

\subsection{Masses, radii and its uncertainties}
For five of our targets, the relative errors in masses, $\Delta M/M$,  
are higher than 1\%. These include the two systems observed with 
Radcliffe/GIRAFFE -- ASAS J010538 
(4\% for both components) and ASAS J174626. (1.8 and 2.2\% for the primary 
and secondary respectively) and three from AAT/UCLES data --  ASAS J014616 
(1.9 and 2.0\%), ASAS J150145 (1.2\% for both) and ASAS J213429 (2.7 and 2.9\%). 
For two targets,we reached the error in masses below 0.2\%. These are
ASAS J053003 (UX~Men, 0.164 and 0.168\%) and ASAS J010934 (AI~Phe) 
for which we reached 0.091 and 0.087\%  precision in the masses of the primary 
and secondary components respectively. It is worth noting that UX~Men and 
AI~Phe were observed spectroscopically only 8 times each. Three other systems with 
high mass precisions are ASAS J085524 (0.22 and 0.24\%, 3 ThAr and 2 iodine
cell calibrated spectra), ASAS J155259 (0.22 and 0.25\%, 4 iodine cell
calibrated spectra) and ASAS J155358 (0.32 and 0.31\%, 4 iodine cell
calibrated spectra). A similar level of precision ($\sim$0.4\%) in $M \sin^3{i}$ 
was reported for only few other eclipsing binaries, like e.g. AD~Boo, VZ~Hya and 
WZ~Oph by \citet{cla08} but in these cases the number of RV measurements for 
each system reached 100. Our results presented in Tab. \ref{tab_master} include 
also an uncertainty in the inclination angle. The relative precision of the 
$M \sin^3{i}$ only is 0.08 \% for our RV dataset of AI~Phe. 

Contrary to the derived masses, our radii estimations are not as precise. We 
were using only the available ASAS photometry which on the one hand provides 
over 8 years of observations and a full phase coverage but on the other hand, 
the total scatter ($\sim6\sigma$) of ASAS data-points often exceeds 0.1 $mag$. 
Also, when eclipses are short-lasting, occupying a small fraction of the 
orbital period, the number of single data-points during an eclipse is low, 
often below 20. This issue has its direct impact on the precision of such 
parameters like the sum and ratio of radii, temperature ratio and the inclination. 
In the worst case -- ASAS J014616 -- our precision in radii is about 30\%. Seven 
other systems have their radii measured with uncertainties between 10 and 20\% for 
the primaries and between 10 and 27\% for the secondaries. However, on the other 
side of the scale there are five systems with radii precision equal to or better 
than 3\%. These include ASAS J010934 (AI~Phe; 2.7 and 2.5\%), ASAS J053003 
(UX~Men; 2.7 and 2.9\%), ASAS J071626 (1.9 and 1.7\%), ASAS J085524 (1.1 and 1.0\%) 
and ASAS J193044 (V415~Aql; 3.0 and 2.8\%). The results given above could be 
significantly improved if one had high precision photometry available. 

\begin{table}
\caption{Phases of the primary and secondary minima of 8 eccentric binaries,
as given by PHOEBE.}\label{tab_min}
\begin{tabular}{crr}
\hline
ASAS ID & $\varphi_{prim}$ & $\varphi_{sec}$ \\
\hline \hline
010934-4615.9 & 0.01815 & 0.47633 \\
014616-0806.8 & 0.00320 & 0.49647 \\
042041-0144.4 & 0.95698 & 0.53672 \\
042724-2756.2 & 0.00247 & 0.49756 \\
071626+0548.8 & 0.96371 & 0.54653 \\
162637-5042.8 & 0.99426 & 0.50569 \\
185512-0333.8 & 0.99759 & 0.50243 \\
213429-0704.6 & 0.02914 & 0.47022 \\
\hline
\end{tabular}
\end{table}

\begin{figure}
\includegraphics[width=\columnwidth]{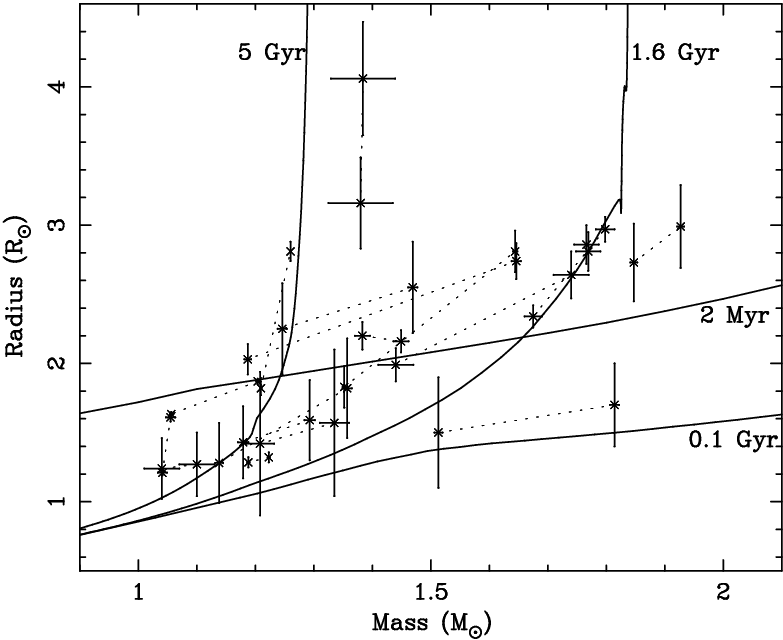}
\caption{Masses and radii of 18 investigated systems with $1\sigma$ uncertainties
overplotted with the $Y^2$ isochrones for solar metallicity and ages 
of 2 Myr, 0.1, 1.6 and 5.0 Gyr. Each isochrone is plotted as a solid line and labeled
with its corresponding age. The components of the same pair are connected with 
thin dotted lines.}\label{fig_mr}
\end{figure}

\subsection{Age estimations}

As pointed out by \citet{bla08}, a 3\% level of precision is already 
suitable for performing reliable tests of the evolutionary models.
Among objects from our sample, one may find several that are
interesting. In Figure \ref{fig_mr} we have shown all our objects in a mass-radius
plane. The components of each binary are connected with a dotted line. 
For a comparison we have also plotted four $Y^2$ isochrones 
\citep{yi01} for the ages of 2 Myr, 0.1, 1.6 and 5 Gyr and solar 
metallicity.

We used the full set of $Y^2$ isochrones of solar metallicity to roughly 
estimate ages of the investigated systems. The isochrones were
fitted by eye to our measuremenst of masses and radii. As
one can see, many systems occupy a region where pre-main-sequence (PMS) 
isochrones (2 Myr as an example) overlap with older ones (1.6 or 5.0 Gyr). 
14 of our systems are in this area and for most of 
them it was not straightforward to determine their evolutionary stage in 
the $M-R$ plane especially given the fairly large uncertainties in radii. 
In order to gain more confidence in the estimated ages, we also verified
whether the resulting temperature ratios for a given isochrone
are consistent with those determined from the light curves. Still
for many of the systems we could find two isochrones that fit
to the systems's parameters. Obviously the estimated ages should be 
treated as very preliminary ones.

Three of the binaries seem to fit better to a PMS isochrone ---
ASAS J042724 (the two solutions are 0.002 and 3.0 Gyr), ASAS J085524 
(0.003 and 7.0 Gyr) and ASAS J162637 (0.0015 and 4.0 Gyr). Four other 
systems, ASAS J174626 (0.002 and 2.0 Gyr), ASAS J193044 (V415 Aql; 0.0015 
and 1.5 Gyr), ASAS J195113 (0.0015 and 3.0 Gyr) and ASAS J213429 (0.008 and
6.0 Gyr), also seem to prefer a PMS isochrone but this is not as clear as for the
previous three. The systems ASAS J014616 (0.006 and 3.0 Gyr), 
ASAS J155259 (0.002 and 2.5 Gyr), ASAS J042041 (0.006 and 4.0 Gyr)
and ASAS 155358 (0.0015 and 1.2 Gyr) seem to fit better to the more evolved 
isochrone. For six other binaries
ASAS J010538 (4.0 Gyr), ASAS J010934 (AI Phe; 5.0 Gyr), ASAS J023631 (3.5
Gyr), ASAS J053003 (UX Men; 2.75 Gyr), ASAS J071626 (8.0 Gyr) and ASAS J185512
(0.4 Gyr) we obtained only one ischrone each consistent with their parameters.
The remaining system, ASAS J150145, is composed of nearly twin stars
so to estimate its age one should use other indicators. It is worth noting that 
the oldest binary, ASAS J071626 (8 Gyr), despite having 
relatively short period ($P\sim 11.5$ d), has a significantly eccentric orbit 
($e\simeq 0.2$). For AI Phe and UX Men our estimates of the age are close to the 
literature values \citep{las02}. 

\subsection{Comparison with literature}
\subsubsection {ASAS J010934-4615.9 = AI Phe}
This system was reported as an eclipsing binary by \citet{str72}
and as a double-lined spectroscopic binary by \citet{imb79} who also
determined the first orbital soultion. This solution was later improved by 
\citet{hri84} who used Imbert's radial velocity data and new
\emph{U, B, V, R, I} photometry to obtain a full set of physical parameters
of the system. Later \citet{and88} used new CORAVEL radial velocity
data together with the unpublished \emph{u, v, b, y} light curves and obtained 
a new model of the system improving parameters' uncertainties by almost an 
order of magnitude. However, the most up-to-date solution was given later 
by \citet{mil92}. They used RV from \citet{and88}, \emph{U, B, V, R, 
I, u, b, v, y} light-curves and far-UV observations in 3 bandpasses of 
the systems' eclipse from \citet{mil81} and processed the data with 
their improved version of the WD code. More recently, a reanalysis of 
the \citet{and88} RV's was obtained by \citet{kar07} with a further
improvement of the spectroscopic parameters. However, their solution can 
be considered as disputable since they seem not to have included 
several uncertainties (of the orbital period, for example) into the total 
error budget and thus presumably underestimated the final errors.

Our solution, based \emph{only} on our RV data and ASAS photometry is
compared with the ones from \citet{and88} and \citet{mil92} in Table 
\ref{com_aiphe}. In general our results are in agreement with the previous 
papers. One should note that having only 8 RV measurements for every component, 
we were able to improve the spectroscopic results of \citet{and88}, who had
46 datapoints for every component, by a factor of 2 to 4. Unfortunately, 
having only one light-curve from ASAS we cannot compete with the results of 
\citet{mil92} who used light curves from 12 bandpasses; most of which
were more precise than ours. Still, their phase coverage is not complete 
and the orbital and physical parameters might be improved with 
high-precision photometry. Our solution converged to a somewhat 
different value of the orbital inclination than Andersen's and Millone's. 
This is of course the reason for the discrepancy in absolute masses between 
the solutions. Our $M\sin^3i$ is however far more precise. Thus we may 
conclude that this systems' parameters can be derived with an unprecedented 
precision (possibly $\sim 0.01$ \% in masses and $\sim 0.1$ \% 
in radii) if only one had more iodine RV measurements and millimagnitude
photometry.

\subsubsection {ASAS J053003-7614.9 = UX Men}
UX Men, as well as AI Phe, was discovered to be a variable in Bamberg 
patrol plates and reported by \citet{str66}. The first orbital solution was 
obtained by \citet{imb74} and the first full physical analysis, based on
Imberts' radial velocities and \emph{u, b, v, y} photometry, was done
by \citet{cla76}. Later, this photometric dataset was reanalised 
together with the new CORAVEL radial velocities by \citet{and89} who 
obtained the most up-to-date solution for UX Men. Comparison of this
solution with our results is shown in Table \ref{com_uxmen}.

Again, our results are in general in agreement with
Andersen's. As for AI Phe, having 8 radial velocity measurements of UX Men,
we succeeded to reach a better precision in the spectroscopic parameters
than Andersen et al. with 29 datapoints for the primary and 31 for the secondary.
In our model we fixed the eccentricity to 0. We found no signifficant 
improvement in the best-fitting model (in terms of $rms$) when $e$ and 
$\omega$ were set as free prameters and the resulting $e$ was undistinguishable 
from 0 within the formal errors. The non-zero eccentricity
may be however induced by a putative third body, found in NACO
images by \citet{tok06} about 0.751 arcsec from the binary. 

\subsubsection {ASAS J193044+1340.3 = V415 Aql}
V415 Aql was first reported as a variable by \citet{hof36}. The first ephemeris
2428670.532 + $E \cdot$ 2.4628 d was determined by \citet{gut39}.
The double-lined character was not previously reported thus there was
no radial velocity curve obtained to date and the only known light curve 
analysis was done by \citet{bra80} who derived an orbital period of
2.46273 d. Our RV data, despite having only 3 measurements for every component,
clearly shows that the actual period is two times longer. This is confirmed by 
the ASAS light curve exhibiting unequal eclipses (see Fig. \ref{fig_master}).
If divided by 2, our period is in a good agreement with the value from 
\citet{bra80}. Unfortunately, the uncertainties of their results are unavailable. 
In Table \ref{com_v415aql} we compare their results with ours. 
The improvement in the derived parameters is obvious. Also the distance
estimate given by \citet[][see Tab. \ref{tab_info} of this paper]{bra80}
should be now treated with caution. 
 
\begin{table}
\caption{Comparison of our results for AI Phe with previous studies.}
\label{com_aiphe}
\begin{tabular}{lccc}
\hline
Parameter& Andersen & Milone &this \\
	 & et al. (1988)& et al. (1992)& paper\\
\hline\hline
$P$ [d]			&\multicolumn{2}{c}{24.592325(8)}& 24.59241(8)\\
$K_1$[km s$^{-1}$]	&\multicolumn{2}{c}{50.95(8)}& 51.16(3)\\
$K_2$[km s$^{-1}$]	&\multicolumn{2}{c}{49.20(8)}& 49.11(2)\\
$v_\gamma$[km s$^{-1}$]&\multicolumn{2}{c}{-1.84(4)}& -0.750(12)\\
$q$			&\multicolumn{2}{c}{1.034(2)}& 1.0418(8)\\
$M_1\sin^3i$ [M$_\odot$]&\multicolumn{2}{c}{1.194(4)}& 1.1922(10)\\
$M_2\sin^3i$ [M$_\odot$]&\multicolumn{2}{c}{1.234(5)}& 1.2421(10)\\
$a$ [AU]		& 0.2225(2) & 0.2223(3) & 0.22371(9)\\
$e$			& 0.1890(67)& 0.1889(5) & 0.187(4)\\
$\omega$ [$^\circ$]	& 109.60(67)& 109.78(3) & 110.1(9)\\
$i$ [$^\circ$]		& 88.45(3) & 88.45(1) & 84.4(5)\\
$M_1$ [M$_\odot$]	& 1.195(4) & 1.190(6) & 1.2095(11)\\
$M_2$ [M$_\odot$]	& 1.236(5) & 1.231(5) & 1.2600(11)\\
$R_1$ [R$_\odot$]	& 1.816(24)& 1.762(7) & 1.82(5) \\
$R_2$ [R$_\odot$]	& 2.930(48)& 2.931(7) & 2.81(7) \\
$T_2/T_1$		& 0.794(27)& 0.816(31)& 0.820(15) \\
\hline
\end{tabular}
\end{table}

\begin{table}
\caption{Comparison of our results for UX Men with previous studies.}
\label{com_uxmen}
\begin{tabular}{lccc}
\hline
Parameter& Andersen & this \\
	 & et al. (1989)&  paper\\
\hline\hline
$P$ [d]			& 4.181100(1) & 4.181096(3)\\
$K_1$[km s$^{-1}$]	& 87.41(25) & 87.31(12)\\
$K_2$[km s$^{-1}$]	& 90.28(17) & 89.90(8)\\
$v_\gamma$[km s$^{-1}$]& 48.47(17) & 48.86(5)\\
$q$			& 0.968(3) & 0.9712(16)\\
$M_1\sin^3i$ [M$_\odot$]& 1.238(6) & 1.2229(15)\\
$M_2\sin^3i$ [M$_\odot$]& 1.198(7) & 1.1878(15)\\
$a$ [R$_\odot$]	& 14.678(25) & 14.639(12)\\
$e\sin{\omega}$	& 0.0025(50) & 0.0(fix)\\
$e\cos{\omega}$	& 0.00083(7) & 0.0(fix)\\
$i$ [$^\circ$]		& 89.6(1) & 89.86(15)\\
$M_1$ [M$_\odot$]	& 1.238(6) & 1.2229(15)\\
$M_2$ [M$_\odot$]	& 1.198(7) & 1.1878(15)\\
$R_1$ [R$_\odot$]	& 1.348(13)& 1.321(36) \\
$R_2$ [R$_\odot$]	& 1.274(13)& 1.285(37) \\
$T_2/T_1$		& 0.993(32)& 0.992(18) \\
\hline
\end{tabular}
\end{table}

\begin{table}
\caption{Comparison of our results for V415 Aql with previous studies.}
\label{com_v415aql}
\begin{tabular}{lccc}
\hline
Parameter& Brancewicz \& & this \\
	 & Dworak (1980)&  paper\\
\hline\hline
$P$ [d]			& 2.462730 & 4.925489(16)\\
$a$ [R$_\odot$]	& 11.60 & 18.44(8)\\
$R_1$ [R$_\odot$]	& 3.26 & 2.97(2) \\
$R_2$ [R$_\odot$]	& 4.05 & 2.82(8) \\
$T_2/T_1$		& 0.918 & 0.974(16) \\
\hline
\end{tabular}
\end{table}

\section{Summary}
Out of the eighteen eclipsing binaries in our sample, 15 are new 
systems discovered by ASAS and fully characterized in this paper 
for the first time with the help of our RV measurements and ASAS
photometry. Even though challenging seeing conditions 
during our observing runs at the AAT/UCLES prevented us from using 
the iodine cell technique for binary stars to the full extent of 
its capabilities \citep{Konacki:05::,Konacki:09::}, 
we were able to derive precise RVs for many 
of the binaries in our sample. In the best cases, when we were able 
to obtain useful spectra with the iodine cell, the precision of mass 
estimation is better than 1\% . In particular in the case of
AI~Phe when we were able to apply the tomographic disentangling
of it composite spectra, the precision in $M \sin^3{i}$ is 0.08\%
with just 8 RV measurements for each component.  This result can be 
improved by an order of magnitude with higher SNR spectra and 
millimagnitude photometry. 

The iodine cell technique for double-lined and eclipsing binary stars 
\citep{Konacki:09::}, is no doubt a powerful tool for stellar astronomy. 
Its applications will provide new challenges not only for the models of 
stellar structure and evolution but also for the techniques used to 
determine stellar metallicity. With masses and radii of stars accurate 
at the level of 0.01\%, it will be challenging to determine stellar 
metallicity with an adequate precision.

\section{Acknowledgments}
We would like to thank Dr. Stephen Marsden from the Anglo-Australian
Observatory and the AAO astronomers for their help during 
our observating runs on the AAT. We would like to thank Dr. David Laney
from the South African Astronomical Observatory for his help during 
our observing runs on the Radcliffe telescope.
 
This work is supported by the Foundation for Polish Science through 
a FOCUS grant and fellowship, by the Polish Ministry of Science and Higher 
Education through grants N203 005 32/0449 and 1P03D-021-29. The observations on the
AAT/UCLES have been funded by the Optical Infrared Coordination network (OPTICON),
a major international collaboration supported by the Research Infrastructures 
Programme of the European Commissions Sixth Framework Programme. This research 
has made use of the Simbad database, operated at CDS, Strasbourg, France.
This publication makes use of data products from the Two Micron All Sky
Survey, which is a joint project of the University of Massachusetts and the
Infrared Processing and Analysis Center/California Institute of Technology,
funded by the National Aeronautics and Space Administration and the National
Science Foundation.

\appendix
\section{Absolute radial velocity measurements}\label{sec_allrv}

\begin{table*}
\caption{Single, absolute RV measurements for all researched objects. 
Time of observation, final error, $O-C$ and calibration 
method are given. ThAr-calibrated spectra are marked with
't', and iodine-calibrated with 'i'.\label{tab_allrv}}
\begin{tabular}{llrrrrrrc}
\hline
ASAS ID & BJD & $v_1$ & $\sigma_{v_1}$ & $O-C_1$ & $v_2$ & $\sigma_{v_2}$ & $O-C_2$ & Calib. \\
 & [2450000. +] & $[$km s$^{-1}]$ & $[$km s$^{-1}]$ & $[$km s$^{-1}]$ & $[$km s$^{-1}]$ & $[$km s$^{-1}]$ & $[$km s$^{-1}]$ & (t/i) \\
\hline \hline
\multicolumn{9}{c}{\it Systems observed with Radcliffe/GIRAFFE}\\
010538-8003.7	& 4007.4184 &  13.767 & 1.358 & -0.369 & -25.896 & 1.955 & -0.244 & t\\
		& 4008.5691 & -48.402 & 1.309 &  0.215 &  34.786 & 2.094 & -2.148 & t\\
		& 4010.5241 & -65.562 & 3.061 &  2.011 &  62.420 & 6.141 &  6.570 & t\\
		& 4166.2696 &  54.048 & 2.553 &  2.579 & -64.431 & 3.172 & -1.571 & t\\

174626-1153.0	& 3903.5470 &  51.205 & 1.810 & -0.647  &-147.011 & 2.830 &  0.258  & t\\
		& 3903.5840 &  57.488 & 2.470 &  2.638  &-151.550 & 3.080 & -1.798  & t\\
		& 3905.3660 &-168.210 & 1.140 &  1.920  &  33.949 & 2.340 & -1.846  & t\\
		& 3906.3570 &  24.759 & 1.410 & -0.074  &-125.532 & 1.430 & -0.507  & t\\
\hline \hline
\multicolumn{9}{c}{\it Systems observed with AAT/UCLES}\\
014616-0806.8	& 4748.1195 & -20.264 & 0.706 & -1.541  &  95.132 & 0.322 &  0.444  & i\\
		& 4748.1301 & -20.239 & 0.719 & -0.813  &  96.055 & 0.326 &  0.584  & t\\
		& 4836.9694 &  -0.589 & 0.718 & -0.688  &  74.374 & 0.346 &  0.674  & t\\
		& 4836.9832 &  -1.255 & 0.706 & -0.236  &  75.848 & 0.327 &  0.901  & i\\
		& 4838.0049 & -41.409 & 0.706 &  1.294  & 119.609 & 0.331 & -1.786  & i\\
		
023631+1208.6	& 4727.1755 &  59.391 & 0.661 & -0.605  & -58.277 & 0.676 & -0.438  & t\\
		& 4748.1731 &  90.371 & 0.576 &  0.218  & -93.688 & 0.453 &  0.076  & t\\
		& 4837.0081 & -59.088 & 1.178 &  0.921  &  86.165 & 0.593 &  1.012  & t\\
		& 4838.0297 &  73.434 & 0.693 &  0.091  & -73.876 & 0.940 & -0.139  & t\\
		& 4839.9661 & -71.914 & 0.815 & -0.525  &  98.097 & 0.619 & -0.627  & t\\

042041-0144.4	& 4727.2386 &  37.134 & 0.333 & -0.149  & -41.185 & 0.295 &  0.485  & i\\
		& 4748.2447 & -52.421 & 0.340 & -0.330  &  56.302 & 0.307 & -0.109  & i\\
		& 4837.0846 &  54.426 & 0.333 & -0.022  & -60.852 & 0.295 & -0.357  & i\\
		& 4838.1014 & -13.658 & 0.334 &  0.453  &  14.537 & 0.301 & -0.188  & i\\
		& 4840.0839 & -62.234 & 0.335 & -0.016  &  67.706 & 0.299 &  0.182  & i\\

042724-2756.2	& 4748.2317 & -46.626 & 0.092 &  0.018  &  92.797 & 0.102 &  0.017  & i\\
		& 4837.0945 & -49.127 & 0.053 &  0.026  &  95.203 & 0.057 &  0.029  & i\\
		& 4838.0716 & -38.469 & 0.047 & -0.082  &  84.879 & 0.049 & -0.008  & i\\
		& 4840.0494 &  53.716 & 0.077 &  0.101  &  -3.094 & 0.081 & -0.091  & i\\
		
071626+0548.8	& 4837.1459 & -29.687 & 0.209 & -0.084  &  60.114 & 0.209 & -0.030  & i\\
		& 4837.1565 & -29.637 & 0.213 &  0.177  &  60.644 & 0.213 &  0.292  & t\\
		& 4838.1523 & -50.603 & 0.212 &  0.180  &  81.076 & 0.212 &  0.419  & i\\
		& 4838.1628 & -51.036 & 0.211 & -0.476  &  80.916 & 0.218 &  0.125  & t\\
		& 4840.1227 & -13.610 & 0.217 & -0.415  &  44.157 & 0.227 &  0.199  & t\\

085524-4411.3	& 4837.1658 &  88.902 & 0.153 & -0.043  & -55.966 & 0.187 & -0.032  & t\\
		& 4837.1788 &  88.908 & 0.106 & -0.119  & -55.867 & 0.161 &  0.162  & i\\
		& 4838.2003 &  67.964 & 0.559 &  0.163  & -31.948 & 0.949 & -0.126  & t\\
		& 4840.1752 & -36.290 & 0.169 &  0.079  &  87.072 & 0.294 &  0.081  & t\\
		& 4840.1872 & -37.096 & 0.134 & -0.326  &  87.617 & 0.279 &  0.168  & i\\

150145-524.2	& 4727.8800 & -22.856 & 0.433 & -0.017  &  31.536 & 0.404 & -0.240  & i\\
		& 4746.8978 & -84.121 & 0.425 &  0.404  &  92.804 & 0.402 &  0.217  & i\\
		& 4747.8818 & -49.147 & 0.417 & -0.470  &  58.758 & 0.392 &  0.130  & i\\
		
155259-6637.8	& 4726.8822 & -28.252 & 0.138 & -0.075  &  43.217 & 0.053 &  0.064  & i\\
		& 4747.8951 & 103.241 & 0.162 & -0.182  & -65.103 & 0.071 &  0.030  & i\\
		& 4748.8874 &  66.360 & 0.148 &  0.076  & -34.587 & 0.058 & -0.040  & i\\
		& 4840.2618 &  99.190 & 0.240 &  0.265  & -61.463 & 0.115 & -0.063  & i\\
		
155358-5553.4	& 4725.8940 & -53.747 & 0.245 & -0.045  &  65.831 & 0.118 & -0.039  & i\\
		& 4726.8693 & -84.018 & 0.123 &  0.342  &  97.428 & 0.058 & -0.394  & i\\
		& 4746.9343 &  89.580 & 0.162 & -0.356  & -83.583 & 0.076 &  0.356  & i\\
		& 4748.8984 & -69.368 & 0.132 & -0.101  &  82.366 & 0.064 &  0.241  & i\\

162637-5024.8	& 4725.9654 &  -5.360 & 0.052 &  0.697  & -55.937 & 0.332 & -0.912  & t\\ 
		& 4726.9048 & -48.709 & 0.054 & -0.006  &  -4.294 & 0.337 &  0.406  & t\\
		& 4727.9052 & -83.688 & 0.046 & -0.206  &  36.225 & 0.366 & -0.115  & t\\
		& 4747.9195 & -62.294 & 0.068 &  0.704  &  11.584 & 0.446 & -0.582  & t\\

185512-0333.8	& 4724.9488 &  21.467 & 0.207 & -0.935  & -63.071 & 0.290 &  1.184  & t\\
		& 4726.9390 & -96.468 & 0.064 & -0.207  &  78.198 & 0.183 &  0.152  & t\\
		& 4746.9608 &  62.064 & 0.087 &  0.320  &-111.946 & 0.122 & -0.544  & t\\

193044+1340.3	& 4726.9783 & -49.255 & 0.155 &  0.277  &  80.940 & 0.084 & -0.094  & t\\ 
		& 4727.9801 & -68.078 & 0.086 &  0.110  & 100.998 & 0.188 & -0.588  & t\\
		& 4747.9736 & -47.881 & 0.216 & -0.203  &  79.187 & 0.324 &  0.144  & t\\

\hline
\end{tabular}
\end{table*}

\begin{table*}
\contcaption{}
\begin{tabular}{llrrrrrrc}
\hline
ASAS ID & BJD & $v_1$ & $\sigma_{v_1}$ & $O-C_1$ & $v_2$ & $\sigma_{v_2}$ & $O-C_2$ & Calib. \\
 & [2450000. +] & $[$km s$^{-1}]$ & $[$km s$^{-1}]$ & $[$km s$^{-1}]$ & $[$km s$^{-1}]$ & $[$km s$^{-1}]$ & $[$km s$^{-1}]$ & (t/i) \\
\hline \hline

195113-2030.2	& 4724.9114 & -74.087 & 0.358 & -0.617  &  82.055 & 0.516 & -0.525  & t\\ 
		& 4725.0293 & -72.957 & 0.367 &  0.001  &  82.638 & 0.461 &  0.761  & t\\
		& 4725.9803 & -44.384 & 0.378 &  0.445  &  42.941 & 0.546 &  0.061  & t\\
		& 4727.9442 &  51.471 & 0.349 & -0.281  & -90.995 & 0.415 &  0.004  & t\\
		& 4727.9567 &  52.053 & 0.169 &  0.003  & -91.450 & 0.429 & -0.040  & i\\
		& 4748.9997 &  50.112 & 0.345 &  0.248  & -88.400 & 0.396 & -0.019  & t\\
		& 4749.0133 &  50.088 & 0.168 & -0.149  & -88.821 & 0.411 &  0.078  & i\\

213429-0704.6	& 4726.0762 &  -8.783 & 2.483 & -3.445  & -37.554 & 1.499 & -0.818  & t\\
		& 4727.0779 & -86.844 & 2.353 &  0.271  &  49.419 & 0.859 &  0.087  & t\\
		& 4748.0581 &  37.655 & 2.359 &  1.261  & -81.116 & 0.860 &  0.172  & t\\
\hline \hline
\multicolumn{9}{c}{\it Systems observed with AAT/UCLES with tomographically disentangled spectra}\\
010934-4615.9	& 4726.1839 & -50.010 & 0.067 &  0.039 &  46.563 & 0.032 & -0.005 & i\\
		& 4727.1199 & -54.702 & 0.065 &  0.008 &  51.034 & 0.030 & -0.012 & i\\
		& 4728.1348 & -54.588 & 0.068 & -0.082 &  50.879 & 0.032 &  0.029 & i\\
		& 4748.1047 & -10.478 & 0.065 &  0.042 &   8.647 & 0.031 &  0.037 & i\\
		& 4749.1254 & -29.290 & 0.068 & -0.095 &  26.506 & 0.036 & -0.034 & i\\
		& 4836.9479 &  30.615 & 0.066 &  0.032 & -30.809 & 0.031 &  0.023 & i\\
		& 4837.9913 &  37.186 & 0.067 & -0.067 & -37.231 & 0.029 &  0.000 & i\\
		& 4839.9391 &  45.639 & 0.064 & -0.051 & -45.336 & 0.029 & -0.009 & i\\

053003-7614.9	& 4726.2312 &  10.474 & 0.186 &  0.089  &  88.834 & 0.132 & -0.112  & i\\
		& 4727.2714 & 126.622 & 0.182 & -0.284  & -31.545 & 0.137 &  0.028  & i\\
		& 4747.2639 &  25.567 & 0.186 & -0.128  &  72.924 & 0.141 & -0.196  & i\\
		& 4748.2774 & 131.771 & 0.201 & -0.388  & -36.657 & 0.130 &  0.492  & i\\
		& 4837.1270 &  75.675 & 0.180 & -0.124  &  21.225 & 0.125 &  0.178  & i\\
		& 4838.1106 & -31.653 & 0.180 & -0.027  & 131.483 & 0.125 & -0.060  & i\\
		& 4838.1334 & -32.599 & 0.181 &  0.104  & 132.385 & 0.125 &  0.391  & i\\
		& 4840.0952 & 122.517 & 0.183 & -0.190  & -26.918 & 0.136 & -0.107  & i\\
\hline
\end{tabular}
\end{table*}

\bsp

\label{lastpage}

\end{document}